# Hierarchical Plant Protein Microcapsules for Hydrophilic and Hydrophobic Cargo molecules

Ngoc-Duy Dinh[a,b], Marc Rodríguez-Garcia[a,d], Zenon Toprakcioglu[a], Yi Shen[a,c] and Tuomas Knowles[a]

[a]Yusuf Hamied Department of Chemistry, University of Cambridge, Cambridge CB2 1EW, UK

[b]Department of Biomedical Engineering, The Chinese University of Hong Kong, Hong Kong SAR

[c]School of Biomedical Engineering, The University of Sydney, NSW, Australia

[d]Xampla Ltd, Cambridge, UK

**Abstract**

Microscale hydrogels comprised of macromolecular networks have increasingly been used for applications involving cell encapsulation, tissue engineering and for the storage and release of active cargo molecules. However, the majority of such microgels are formed from nonbiodegradable synthetic polymers, involving harmful solvents, or using animal proteins, such as silk and gelatin, which can have a negative environmental impact and lack sustainability. Furthermore, most encapsulation techniques involve either protecting hydrophobic or hydrophilic cargo, but rarely both. In order to address these issues, we employed droplet-microfluidics to develop novel, plant-derived biodegradable microcapsules capable of containing both hydrophilic and hydrophobic cargo molecules. The microcapsule structure and cargo release rates were controlled by balancing osmotic pressures between the outer and inner phases of the capsules. Moreover, the digestibility of the microcapsules was comparable with that of pure pea protein, thereby enabling the use of these microcapsules for food and beverage applications. In addition, digestive enzymes can trigger the release of the encapsulated active ingredients, and hence, these microcapsules are well suited for the controlled delivery of active nutraceutical or pharmaceutical ingredients. Finally, we investigated the biodegradability of the microcapsules. It was determined that the plant protein microcapsules exhibited 98.0% biodegradability (as compared with cellulose), thereby fulfilling the biodegradability standards stipulated by the International Organization for Standardization (ISO 14851) for microplastics in freshwater conditions (90%). Furthermore, by encapsulating iron, an essential micronutrient within the microgels, we prevent its oxidation and thus show the potential of the capsules in the food and beverage industries. Hence, these microcapsules provide an avenue for sustainable and environmentally friendly encapsulation of both hydrophilic and hydrophobic active ingredients that can be employed in home and personal care products and can be used instead of synthetic polymers to reduce microplastic pollution.

**Introduction**

Microscale hydrogels formed of dense macromolecular networks are extremely important for a variety of applications such as cell encapsulation, tissue engineering and for regulating release of cargo molecules in a controlled manner. Microcapsules are usually comprised of two parts; a core (liquid, gas, or matrix polymer) where the active ingredient are typically encapsulated within, and the shell, which is usually solid and is used to protect and regulate diffusion of cargo molecules (1-5). The shell creates a physical barrier between the external environment and the core, thereby protecting the sensitive active ingredients (flavors, polyunsaturated oils, antioxidants, vitamins, drugs, etc.) from external factors, particularly moisture, ultraviolet (UV), pH, and oxidation. The utility of microcapsules as vehicles for the targeted release and delivery of stabilized active ingredients is considerably important in the biomedical, pharmaceutical, food, cosmetic, personal care, textile, and agricultural industries, amongst others. However, most microcapsules used in these industries are environmentally unfriendly, and are comprised of nonbiodegradable polymers, including polyurethane, poly(melamine-formaldehyde), polyethylene, polystyrene, or polypropylene. Moreover, the breakdown of microcapsules in consumer products due to friction produces microplastics (6-8). Subsequently, these microplastics leach into the wastewater disposal system and enter rivers and the ocean, being ingested by marine life before ending up in our food chain. In an effort to alleviate microplastic pollution, the governments of Britain, the European Union, Canada, the United States, New Zealand, and South Korea have banned the use of synthetic, nonbiodegradable polymers in certain types of personal care and cosmetic products, such as shower and bath gels, hair products, lotions and creams, sunscreens and tanning creams, makeup, perfumes, soaps, exfoliants, toothpastes, cleansing products, hand sanitizers, and shampoos (9-12). Consequently, to fulfill the industry demand, there has been an increased interest in developing biodegradable microcapsules by using materials such as polylactic acid, poly (lactide-co-glycolide), and polycaprolactone. However, these microcapsules are not completely biodegradable, with their fabrication involving harmful solvents (13,14). Nevertheless, biodegradable protein-based microcapsules have been formed from natural, animal-derived sources, such as silk (15-18), lysozyme (19), gelatin (20), and chitosan (21), via controlled self-assembly. However, their application has been challenging owing to production costs, potential allergenicity, and the environmental impact originating from animal-derived feedstocks (22). Plant-derived polysaccharides and polyesters require covalent cross-linking that may sometimes damage the encapsulated active material and inhibit the thorough removal of the unreacted species (23, 24). Plant proteins have been used as shell-forming materials for microencapsulation because they are less allergenic than animal-derived proteins and adhere to the present "green" trend prevalent in the pharmaceutical, cosmetic, and food industries (25). However, most plant-based proteins are poorly soluble in water, posing fundamental challenges in controlling their self-assembly into desired structures. Recently, a method to dissolve plant proteins in water to enable controlled film formation has been developed with applications in food packing and single-use replacement plastics (26).

Conventional microcapsule fabrication methods involve emulsion polymerization, dispersion polymerization, and spray drying. However, they pose challenges, such as enlarged polydispersity, diminished reproducibility, limited functionality, and reduced tunable morphology. To overcome these challenges, various alternative fabrication methods, such as droplet microfluidics, flow lithography microfluidics, electrohydrodynamic cojetting, photolithography, soft lithography-based imprinting, and micromolding, have been developed (27-29). Recently, droplet microfluidics has been reported as the most effective alternative method for forming microcapsules, as it offers exquisite control over multiple fluids at the microscale (30). Monodisperse droplets are generated by the intersection of two immiscible fluids, typically an oil and an aqueous phase. Moreover, the integration of micron-sized junctions with appropriate surface wettability treatments can result in the production of double emulsions with tunable internal structures, which can be used as templates to fabricate microcapsules with desirable features. An enormous range of microcapsules has been fabricated using droplet microfluidics for diverse applications, such as in tissue engineering, drug delivery, cosmetics, and pesticides (30-32). To date, most of the microcapsules that have been generated via droplet microfluidics are comprised of synthetic nonbiodegradable polymers, which usually involve harmful solvents (33-35), animal proteins (36) such as silk (37) and gelatin (38), or plant alternatives derived polysaccharides, such as alginate, which include chemical cross-linking and are thus not completely biodegradable (39, 40).

Here we present a droplet-microfluidic approach for generating plant-based protein microcapsules using tandem emulsification. We demonstrate that the plant protein microcapsules could be used to encapsulate both water-soluble components (hydrophilic), such as vitamin C, vitamin B2 (riboflavin), vitamin B5, vitamin B12 and iron and fat-soluble ingredients (hydrophobic), including fragrances, vitamin D, vitamin E and essential oils. Calcium chloride ($CaCl_2$) was used to balance the osmotic pressure and optimize microcapsule structure and molecular release. The microcapsule shell inhibited interactions between the encapsulated iron and oxidizing chemicals present in food. Furthermore, digestibility experiments show that the plant-based protein hydrogel used in the microcapsules exhibited digestibility comparable with that of pure pea protein, and thus, these microcapsules can be applied in food and beverage products. A two-stage *in vitro* digestibility experiment also confirmed that digestive enzymes (simulated gastric fluid, SGF) triggered the release of cargo. Thus, these microcapsules can be exploited for the controlled delivery of active nutraceutical or pharmaceutical ingredients. The relative biodegradability of the fabricated microcapsules was 98.0% compared with cellulose. Moreover, the microcapsules met the biodegradability requirement mandated by the International Organization for Standardization (ISO 14851) for microplastics in freshwater conditions (90%) (41). Hence, these plant protein microcapsules can be used as environmentally friendly alternatives in various consumer products, particularly in home and personal care products, to reduce microplastic pollution.

**Results and Discussion**

In this study, we develop a tandem emulsification approach involving microfluidics that utilized molecular self-assembly, to engineer the nanoarchitectures of plant proteins in order to form microcapsules. The tandem emulsification device design is shown in Fig. 1. The microfluidic system constituted of two connected microfluidic chips whose surfaces were separately modified prior to droplet formation. The surface functionalization of Chip 1 (Movie S1) was achieved via plasma activation to induce a hydrophilic channel, while that of Chip 2 (Movie S2) was achieved via silanization to induce a hydrophobic channel. Each chip comprised of two inlets and one outlet. The outlet of Chip 1 was connected directly to one of the two inlets of Chip 2, and it comprised the dispersed phase in Chip 2. For the tandem emulsification setup, the three inlets were controlled by a microfluidic flow controller. Hydrophilic cargos (vitamin C, iron) were encapsulated using a triple emulsion system. Vitamin C release was modeled using a fluorescent dye, fluorescein. The fluorescein solution was sonicated in refined coconut oil along with a food-grade surfactant, polyglycerol polyricinoleate (PGPR), to create the primary water-in-oil (W/O) emulsion. Double emulsions were formed in Chip 1 and subsequently transferred to Chip 2, wherein oil-in-water droplets were encapsulated by the continuous oil phase, yielding a triple emulsion, water-in-oil-in-water-in-oil (W/O/W/O). Numerous critical parameters influence the number of internal aqueous core droplets and the sizes of core and shell, which are determined by the flow rate ratio of the inner and middle phases. In flow-focusing microfluidics, the droplet size depends on the flow rates of the dispersed and continuous phases. The breakdown of these droplets is controlled by the viscous shear force of the continuous phase, which stretches the liquid thread downstream, and the surface tension, which holds it together (42). Hence, to achieve a multicore double emulsion, we optimized the ratio of the inner phase flow rate to the middle phase flow rate in the first device and to the outer phase in the second device. The flow rates were controlled by a flow controller with a pressure control of 200/250/400 Pa.

In order to form microcapsules, we utilized the propensity of proteins to undergo self-assembly and form fibrillar-like hydrogels. The self-assembly process by which plant protein microcapsule formation was achieved is shown in Fig 2. Briefly, gelation provides a driving force to unfold protein architectures, via mechanisms such as thermal unfolding and the association-dissociation of protein subunits with subsequent aggregation involving, at least partially, sulfhydryl-disulfide conversion, to form a three-dimensional (3D) gel network. Gelation involved partial protein denaturation, during which relevant intermolecular/intramolecular interactions stabilize native architecture. During microcapsule formation, environmentally friendly acetic acid-water binary mixtures and ultrasound sonication were used to dissolve the proteins in the solution. The solution was prepared by dispersing soy protein isolate (SPI) or pea protein isolate (PPI) at a concentration of 10% (w/v) protein in 42% (v/v) aqueous acetic acid. The subsequent ultrasonication and heating at 90°C for 40 min produced a translucent aqueous solution (Fig. S1). SPI and PPI are poorly soluble in water because they contain a high percentage of intermolecular β-sheets (49%) (26). Sonication and high temperatures denaturize and hydrolyze the proteins into oligopeptides, which subsequently assemble into fibrils of aligned β-sheets. The aggregation of protein molecules via the "hydrophobic effect" and "association phenomenon" yield structured fibrillar networks. Moreover, sonication and high temperatures reduce the proportion of α helices, improving the propensity for β-sheet development in proteins. Notably, β-sheets aggregate in a rigid fashion to yield characteristic structures while assembling into protein fibrils. Thus, they are available to form new structures. Hence, to this effect, sonication and high temperatures were employed to enhance intermolecular interactions via hydrogen bond formation, thereby facilitating β-sheet self-assembly. The solution was then cooled to induce the self-assembly of proteins into intermolecular β-sheet–rich fibrillar aggregates, stabilized by stronger hydrogen bonding to form a hydrogel. This solution was then fed into the microfluidic chip before forming the protein microcapsules. To enhance protein gelation and cargo retention, the microcapsules were rinsed twice with $CaCl_2$ before redispersing in phosphate-buffered saline or deionized water for use. Adding salt ($CaCl_2$) allowed for calcium ions ($Ca^{2+}$) to further promote gelation and increase the 3D network of the hydrogel, resulting in augmented G-values (43). The shell of the microcapsules exhibited strength and stability under shear conditions, such as washing, shaking, or sterilization. Furthermore, washing the microcapsules with $CaCl_2$ helped balance the osmotic pressure. Particularly, in w/o/w emulsions, the balance in the osmotic pressures of the inner and outer aqueous phases is imperative for the cargo retention capacity of the microcapsules.

Osmotic pressure differences also exacerbate undesired leakage through the shell when the concentration of active cargo is in a gradient. Osmotic pressure imposes tensile stress on the shell surface, increasing the rate at which active compounds diffuse out of the capsules and reducing encapsulation efficiency. Thus, effective strategies to avoid leakage are essential for the application feasibility of the microcapsules. To overcome this challenge, we added $CaCl_2$ to the internal phase, commonly used to prevent instability due to water transport to the external phase because of differences in Laplace pressure. In the absence of $CaCl_2$ in the internal phase, primary emulsions are unstable after 1 day of capsule generation, with complete molecular release of fluorescein being observed after 1 month of formation, as shown in Fig. 3A–C.

Conversely, in the presence of $CaCl_2$ solution, primary emulsions remained stable (black) after 1 day of formation and fluorescein release was not observed even after 1 month of observation, as shown in Fig. 3D–F. Preparing the primary solution via sonication is another effective strategy to enhance emulsion stability and can be used to fabricate smaller emulsions than that achieved just by vortexing, as shown in Fig. S2. Moreover, the fluorescein intensity of sonicated versus vortexed samples was monitored using fluorescence intensity measurements. It was determined that when the primary emulsion is generated using vortexing rather than sonication, the emulsion is more stable (Fig. 3G). A $CaCl_2$ concentration dependence was observed. Moreover, the fluorescein intensity of the primary emulsion in the absence of $CaCl_2$ addition was 3.6 times lower than when $CaCl_2$ was added (Fig. 3F). Accordingly, the $CaCl_2$ concentration was optimized at 0.5 M and the microcapsules were de-emulsified and washed with $CaCl_2$ to enhance the effect of $Ca^{2+}$ on the 3D network gels.

The addition of $Ca^{2+}$ facilitated the formation of a stable 3D gel network and accomplished electrostatic shielding, ion-specific hydrophobic interactions, and cross-linking of adjacent anionic groups via the formation of protein-$Ca^{2+}$-protein bridges (44). Hence, washing the microcapsules with $CaCl_2$ substantially enhanced the elastic modulus and textural properties of the hydrogels, potentially facilitated by a $Ca^{2+}$ bridging mechanism, and generated a well-structured composite hydrogel. The gel hardness increased optimally with 0.5 M $CaCl_2$ (Fig. 4A & D). Furthermore, the microcapsules did not swell, break, or release core encapsulated material at the above mentioned $CaCl_2$ concentration. Conversely, the microcapsules that were not washed with $CaCl_2$ swelled and broke because of the osmotic pressure imbalance, releasing the encapsulated core and leading to the loss of protection and regulation of the active ingredients (Fig. 4B, C, E, & F). The addition of the optimized concentration of $Ca^{2+}$ (0.5 M) promoted protein-protein interactions, rendering a more ordered and compact protein network structure with improved hardness of the composite hydrogel filled with modified protein microcapsules.

The multifunctionality of the encapsulation method presented in this study was demonstrated by fabricating particle/oil/particle microcapsules and encapsulating crystalized vitamins in oil, as shown in Fig. 5A. The tandem emulsification system that was used is shown in Fig. 5B. Food-grade agarose microparticles were formed via microfluidics (Movie S3). Hydrophilic active ingredients, such as fluorescein, vitamin C, vitamin B2, and iron were encapsulated within agarose microparticles. The middle phase comprised of a coconut oil and food-grade surfactant PGPR mixture. Agarose in coconut oil emulsions were thus made (Chip 1) before being transferred to Chip 2. The agarose microparticles in the oil were encapsulated in the protein phase (Chip 2). The final protein microcapsules are shown in Fig. 5C and Fig. S2. This technique enabled the encapsulation of water-soluble active ingredients or drugs in a protein microcapsule, which can be suitable for applications involving the food, micronutrient, and pharmaceutical industries.

Many micronutrients, such as vitamin C and iron, are sensitive to UV light and oxidizing chemicals, which degrade or change their oxidative states, thereby limiting their absorption after ingestion. Therefore, encapsulation can improve micronutrient stability against these challenges for both individual and co-encapsulated formulations. Furthermore, the protein microcapsules that address these stability challenges can facilitate the implementation of staple food fortification to support global health initiatives for treating micronutrient deficiencies. In this study, we also show that these protein microcapsules could be fabricated by encapsulating fat-soluble ingredients, such as fragrances, essential oils, and vitamin E (Fig. 5D, E), which possess a wide range of applications in cosmetic and personal care products. Furthermore, the relative degree of degradation of the plant protein microcapsules was 98.0%, as compared with cellulose, which fulfilled the biodegradability requirement of microplastics in freshwater conditions (90%), as stipulated by ISO 14851 (Tables S1, S2, and Fig. S3). Hence, plant protein microcapsules can be used in place of microcapsules comprised of synthetic polymers, in cosmetic and personal care products to reduce microplastic pollution.

To demonstrate the application of microcapsules in the long-term protection of active ingredients, we assessed the ability of the microcapsules in encapsulating iron, as iron is an essential WHO-recommended micronutrient, which requires fortification in cereals, dairy products, and juices. However, iron fortification often causes strong color and taste changes in food products (45). Furthermore, polyphenols present in food catalyze iron oxidation from a highly bioavailable ferrous ($Fe^{2+}$) state to a ferric ($Fe^{3+}$) state, which exhibits poor bioavailability (46). We examined whether microcapsule encapsulation prevents interactions between the encapsulated iron and oxidizing chemicals present in food by adding microcapsule-encapsulated and -unencapsulated iron to polyphenol-rich banana milk. The samples were then analyzed for color change after 1 min and again after 2 weeks. The banana milk containing microcapsule-encapsulated iron, demonstrated no color change after 2 weeks

(tube 4), similar to the control (tube 1). Conversely, unencapsulated iron interacted with the polyphenols in the banana milk, producing a purple-blue color (tubes 2-3) These results are summarized in Fig. 6A, B. The reducing effect and high bioavailability make the novel microcapsule model promising for nutritional applications.

We then sought to determine whether the reducing effect of plant proteins on iron could reduce this micronutrient to its more bioavailable Fe(II) state. In a protein microcapsule encapsulating ferric chloride ($FeCl_3$) solution, Fe(III) ions were converted to Fe(II) ions. The addition of increasing amounts of protein fibrils to an $FeCl_3$ solution afforded higher amounts of Fe(II) ions (Fig. 6C, D). The protein microcapsules encapsulating iron exhibited a greater Fe(II) ion concentration than β-lactoglobulin (BLG) fibrils: $FeCl_3$ and protein microcapsules without encapsulated iron. The concentration of Fe(II) ions was determined by absorbance of Fe(II)-phenanthroline complex measured at 512 nm, as shown in the images to the right of the plot, where greater color intensity corresponds to a significant conversion of Fe(III) ions to Fe(II) ions (Fig. 6). The stabilization of highly reactive iron salts is crucial in micronutrient fortification because of the risk of color loss, metallic taste, and, eventually, rancidity.

The measured digestibility of the microcapsule was 100.8% ± 2.0%, while the control sample (PPI, Cambridge Commodities Ltd) demonstrated a Boisen protein digestibility of 99.7% ± 2.0%. Therefore, plant protein microcapsules exhibit a digestibility comparable with pure pea protein with promising applications in food and beverage products. Controlled release experiments were performed using digestive enzymes (SGF) via a two-stage *in vitro* digestibility test (Fig. S4). The results showed that the microcapsules can be used as ingestible micronutrient delivery platforms with the potential to improve micronutrient deficiency in developing countries. This approach could also be applied to microencapsulation and oral delivery in the food and nutraceutical industries.

In conclusion, we have demonstrated the use of microfluidics to form hierarchical plant-based protein microcapsules using tandem emulsification via flow focusing. By utilizing the propensity of proteins to undergo self-assembly into structures that mimic hydrogels, we were able to generate micron-sized hydrogels which could protect the active ingredients. The structure and retention of the active ingredients were optimized with $CaCl_2$ to balance the osmotic pressure. The microcapsules prevented the oxidation of encapsulated iron in food for up to 2 weeks. We also found that plant protein microcapsules exhibited a digestibility comparable with pure pea protein. Hence, the microcapsules presented in this study can potentially be applied in food and beverage products. Furthermore, the microcapsules demonstrate controlled release on being triggered by digestive enzymes (SGF), suggesting that they can be exploited for the controlled delivery of active nutraceutical or pharmaceutical ingredients. Finally, the relative biodegradation of the microcapsules was 98.0% compared with cellulose, thus meeting the 90% biodegradability requirement stipulated by ISO 14851. Hence, these microcapsules can be used for environmentally friendly encapsulation in home and personal care products to reduce microplastic pollution. The plant protein microcapsules can have numerous applications in the food, nutraceutical, pharmaceutical, cosmetic, personal care, and agriculture industries.

**Materials and Methods**

**Fabrication of the tandem emulsification system**

Based on our previous studies, we employed tandem emulsification using flow-focusing polydimethylsiloxane (PDMS) microfluidic devices fabricated via soft lithography (Fig. 1). The PDMS prepolymer was cast and cured on SU-8 3050 and SU-8 3025 molds following a standard protocol (47-49). The fabrication of the tandem emulsification system was based on the 3D instead of the 2D flow-focusing microfluidic device, as in the previous study (50). Briefly, the multistep photolithographic process involved the fabrication of a 25-μm photoresist

SU-8 3025 layer in the first master (Chip 1) for the inner channel, followed by the fabrication of a 50-μm layer for the outer channel. In the second master (Chip 2), a 100-μm photoresist SU-8 3050 layer was fabricated for the inner channel, followed by a 150-μm layer for the outer channel. To fabricate the PDMS microfluidic chips, two different PDMS slabs were aligned and bonded by plasma oxidation. Water or ethanol droplets were added between the two PDMS slabs after plasma oxidation to align the microfluidic channels. The chips were placed in an oven at 65°C overnight to complete the bonding between the two layers. The device comprised two connected microfluidic chips whose surfaces were modified separately before the connection. The surface functionalization of Chip 1 (Video S1) was achieved via plasma activation to induce a hydrophilic channel, while that of Chip 2 (Video S2) was achieved via silanization to induce a hydrophobic channel. Each chip comprised two inlets and one outlet: the outlet of Chip 1 was connected directly to the inlet of the dispersed phase in Chip 2. The three inlets were controlled using vacuum pumps for the tandem emulsification setup.

**Microcapsule fabrication**

PPI (80% protein) and SPI (90% protein) were purchased from Cambridge Commodities Ltd. SPI (625 mg) was dispersed in 5 mL of 42% v/v aqueous acetic acid (Sigma-Aldrich) and shaken well until a turbid and highly viscous dispersion was obtained. For protein solubilization, the mixture was sonicated using an ultrasonic homogenizer (Bandelin, HD2070, 70 W) for 40 min (amplitude, 40%; on-time pulse duration, 0.7 s; off-time pulse duration, 0.3 s). After 30–45 min, a completely translucent liquid solution was obtained.

Primary emulsions were prepared by sonicating 200 μL fluorescein (150 mg/mL) in 2 mL refined coconut oil with 3% v/v surfactant PGPR (Danisco). The primary emulsion (inner phase), PPI solution (middle phase), and 3M Novec 7500 oil (2% v/v 008-Fluorosurfactant, outer phase) were pumped into the microfluidic chips using a pressure-driven system (Elveflow

OB1). Different pressure rates were tested until uniform and continuous generation of a monodisperse population of microcapsules with ~300-μm diameter was achieved (pressures: inner phase, 200 mbar; middle phase, 400 mbar; outer phase, 200 mbar).

**Microcapsule-encapsulated cargo**

The microcapsule encapsulated both water- and oil (fat)-soluble ingredients. The microcapsule-encapsulated water-soluble ingredients were fabricated using triple emulsion. The water-soluble ingredients, including vitamin C, vitamin $B_2$ (riboflavin), vitamin $B_5$, vitamin $B_{12}$, and iron, were encapsulated in primary emulsion plant-based oils with oil-soluble PGPR surfactant using sonication. The presence of an oil-soluble surfactant provided a more stable primary emulsion. The concentration of the oil-soluble surfactant in the lipophilic phase was in the range of 2%-4% w/w. Then, primary emulsions were encapsulated in the plant protein phase. To stabilize the hydrophilic phase, pectin was used as a rheology modifier (e.g., a hydrocolloid) for mixing with the aqueous phase, comprising 1% (w/w) high methoxyl pectin solution with a suspension of a soluble active ingredient (riboflavin). The microcapsule-encapsulated oil-soluble ingredients, such as fragrances, vitamin D, and essential oils, were fabricated using double emulsion.

**Iron encapsulation**

The color change in a food matrix after adding iron microcapsules was measured using banana milk slurry, which has a polyphenol-rich food matrix (45, 46). Fortified banana milk was prepared at a concentration of 60 ppm iron in 70 g banana milk (180 g fresh banana with 520 g organic whole milk [3.9% fat, homogenized, pasteurized]). Microcapsule-encapsulated and unencapsulated iron were added to the banana milk, and the color change was quantified over time (after 2 weeks) to assess whether microcapsule encapsulation prevented interactions between the encapsulated iron and oxidizing chemicals present in food. The polyphenols in banana milk catalyzed iron oxidation, thereby changing color from a highly bioavailable ferrous ($Fe^{2+}$) state to a ferric ($Fe^{3+}$) state that exhibited poor bioavailability.

**Measurement of the reducing effect**

BLG (>98%, Davisco Foods International) was used after dialysis, and BLG amyloid fibrils were prepared according to the previous protocol (51). BLG fibrils and iron nanoparticles (Fe-FibBLG) were reacted by *in situ* chemical reduction of $FeCl_3 \cdot 6H_2O$ (97%, Sigma-Aldrich). Fe(III) ions binding BLG fibrils were then chemically reduced by $NaBH_4$ (≥99%, Sigma-Aldrich).

The presence of Fe(II) was measured by adding 1,10-phenanthroline (≥99%, Sigma-Aldrich) to the solution. The distinctive orange color was measured at 512 nm (Tecan, 200 PRO multimode reader). Protein microcapsule-encapsulated Fe(III) showed greater absorbance than BLG fibrils:$FeCl_3$ and protein microcapsules without encapsulated iron.

**Digestibility of plant protein microcapsules**

The digestibility of the plant protein microcapsules was characterized by following the Boisen protocol (52). The digestibility of measured protein was 100.8% ± 2.0%, while the control sample (PPI, Cambridge Commodities Ltd) exhibited a Boisen protein digestibility of 99.7% ± 2.0%.

**Controlled release of microcapsules**

Microcapsule release was controlled by triggering digestive enzymes via a two-stage *in vitro* digestibility test. Pepsin (8 mg) in SGF electrolyte solution was prepared. SGF was prepared by dissolving 0.257 g KCl, 0.061 g $KH_2PO_4$, 1.05 g $NaHCO_3$, 1.38 g NaCl, 0.122 g $MgCl_2(H_2O)_6$, and 0.024 g $(NH4)_2CO_3$ in 1 L deionized water. The pepsin was dissolved in SGF and adjusted to pH 2 by adding a small amount of 1 M HCl. A mixture of 50 mL microcapsule and 200 mL SGF was centrifuged at 37°C and 300 rpm for 60 min and adjusted to pH 2 by adding a small amount of 1 M HCl. The simulated intestinal phase was prepared by dissolving 3 mg pancreatin in 1 ml simulated intestinal fluid (SIF) electrolyte solution (0.253 g KCl, 0.054 g $KH_2PO_4$, 3.57 g $NaHCO_3$, 1.12 g NaCl, 0.335 g $MgCl_2(H_2O)_6$, 0.44 g $CaCl_2 \cdot 2 H_2O$, and 0.23 g bile extract in 1 L deionized water). Then, 400 mL SIF was added to the previous microcapsule in the SGF solution, the sample was incubated at 37°C for 120 min, and the pH was adjusted to 7 by adding a small amount of 1 M NaOH

**Biodegradability of plant protein microcapsules**

The microcapsules were dried for 3-4 h at 85°C and then subjected to an aqueous aerobic biodegradability test in fresh water according to ISO 14851, using cellulose as the reference standard. The ratio of the biochemical oxygen demand (corrected for the control) to the theoretical oxygen demand delineates $O_2$ consumption or the amount of biodegradation (22). After 28 days, as measured using $O_2$ consumption, the biodegradation of the reference sample cellulose was 82.4%, while that of the tested microcapsules was 87.5% (Tables S1 and S2). The extent of biodegradation, as measured by the amount of $CO_2$ produced, was based on the percentage of solid carbon of the test material. The $CO_2$ production of the reference sample cellulose and the microcapsules was 84.2% and 82.4%, respectively, after 28 days. The relative biodegradation of the tested microcapsule was 98.0% compared with cellulose, exceeding the ISO 14851 standard for biodegradability (90%).

**Keywords:**

Plant Protein Microcapsules, Self-assembly Plant Proteins, Microcapsules, Microfluidics

**Competing interests:**

The work described in this paper has been the subject of a patent application filed. T.P.J.K. is a member of the board of directors of Xampla, Ltd, a Cambridge University spin-off company. The remaining authors declare no competing interests.

**Acknowledgements:**

The work was supported by the European Research Council under the European Union's Seventh Framework Programme (FP7/2007-2013) through the ERC grant PhysProt (agreement no. 337969) and ProtCap (agreement no 825803).

**Author contributions**:

N.D.D and M.R.G designed research; N.D.D and M.R.G performed research; Y.S. carried out the reducing effect experiments.; N.D.D wrote the paper; Z.T revised manuscript; T.P.J.K. supervised the research and revised manuscript.

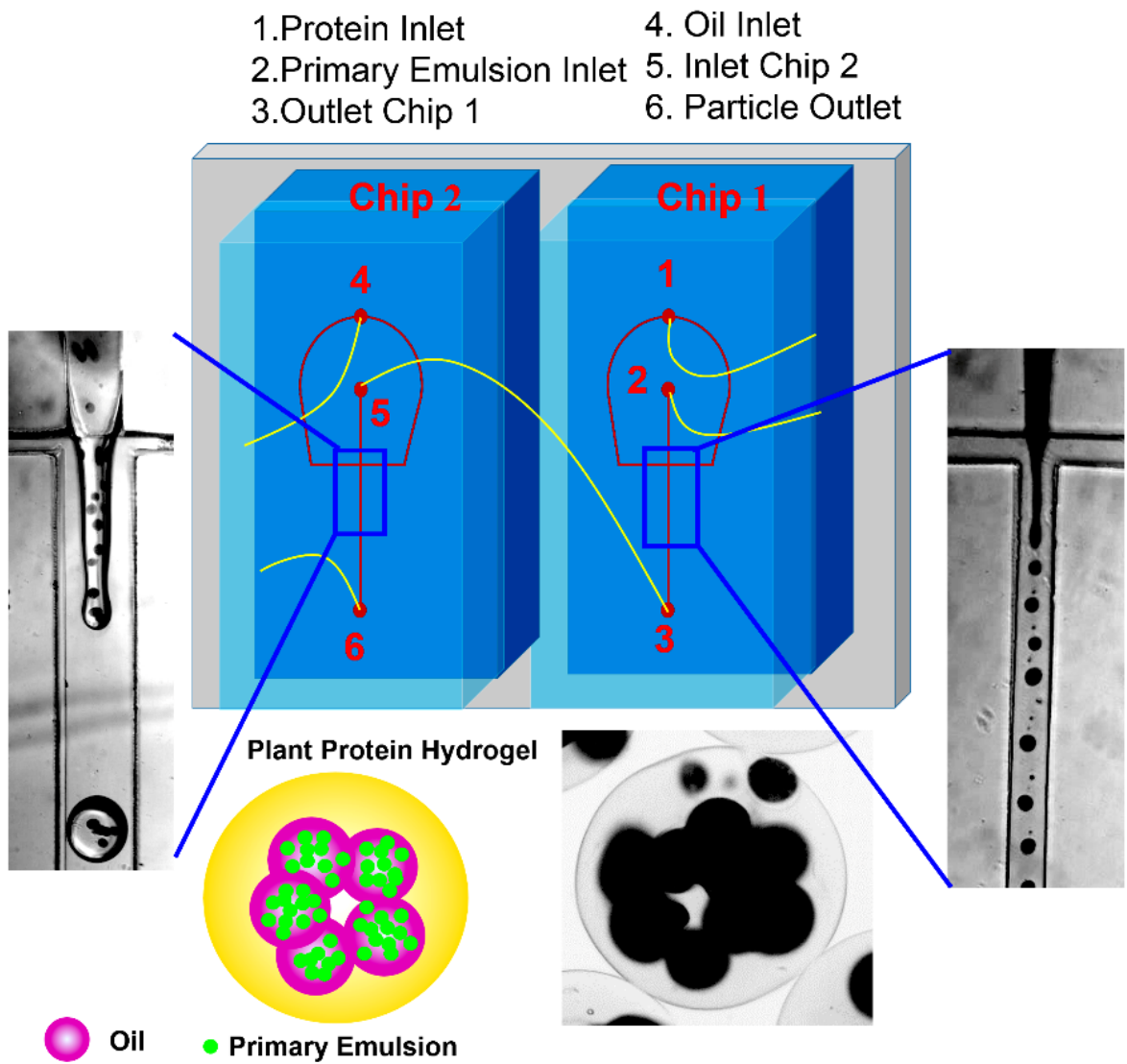

**Fig. 1 Illustration of the tandem emulsification systems for fabricating plant protein microcapsules**. The system is comprised of two microfluidic chips. The surface functionalization of Chip 1 was achieved by plasma activation to induce a hydrophilic channel, whereas that of Chip 2 was achieved by silanization to induce the hydrophobicity within the device. Outlet (3) of Chip 1 is connected directly to inlet (4) of the dispersed phase in Chip 2. Double emulsions were formed in Chip 1 and continuously transferred to Chip 2, wherein oil-in-water droplets are encapsulated by the continuous oil phase resulting in formation of triple emulsions: water-in-oil-in-water-in-oil (W/O/W/O). The three inlets were controlled by a microfluidic flow controller with a pressure control of 200/250/400 Pa.

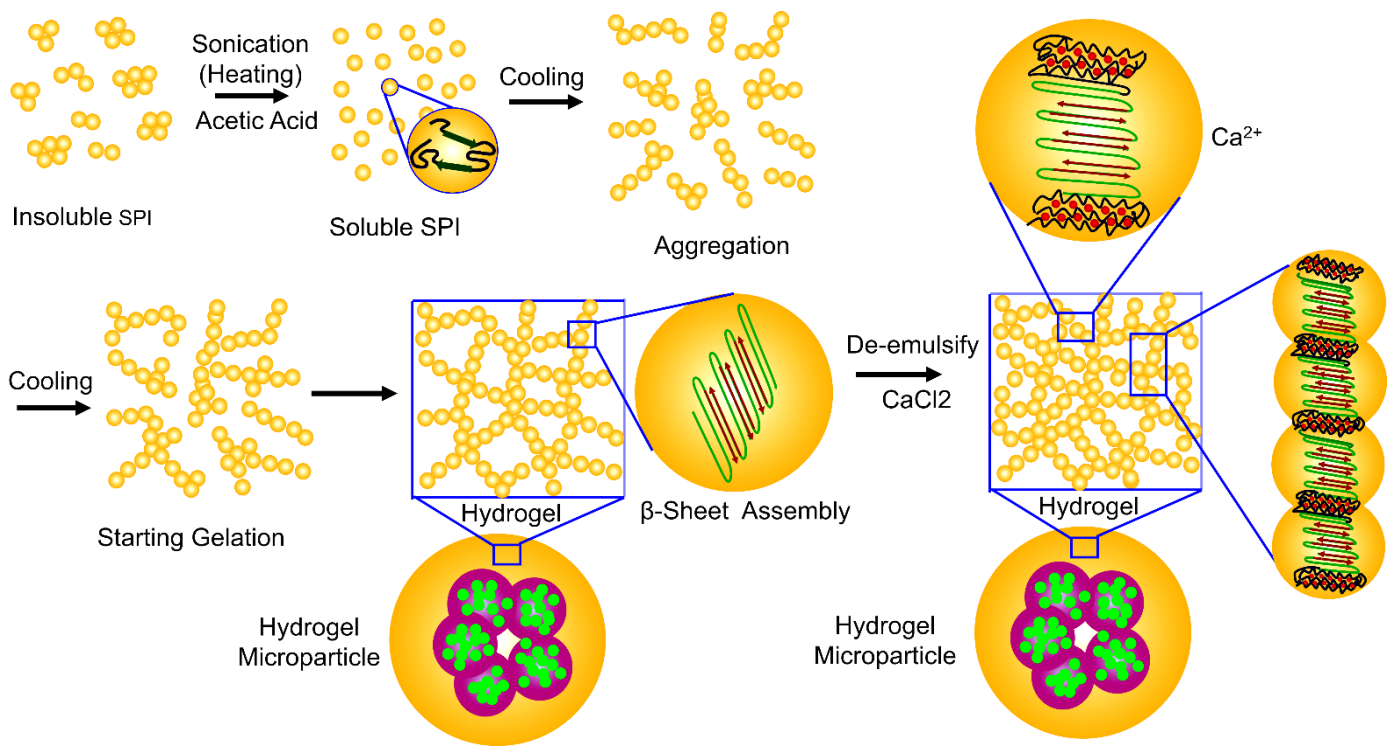

**Fig. 2 Schematic representation of the self-assembly process resulting in the formation of plant protein microcapsules.** The solution was prepared by dispersing soy protein isolate (SPI) or pea protein isolate (PPI) at 10% w/v protein concentration in a 42% v/v aqueous acetic acid solution. The solution was then ultrasonicated and heated (90°C) for 40 min, producing a translucent aqueous solution. The solution was then cooled to induce protein self-assembly into intermolecular β-sheet–rich fibrillar aggregates, stabilized by hydrogen bonding and resulting in a hydrogel. This solution was fed into the microfluidic device and the microcapsules were subsequently rinsed twice with calcium chloride to enhance the gelation and mechanical structure of the hydrogel before being redispersed in phosphate-buffered solution or deionized water for use.

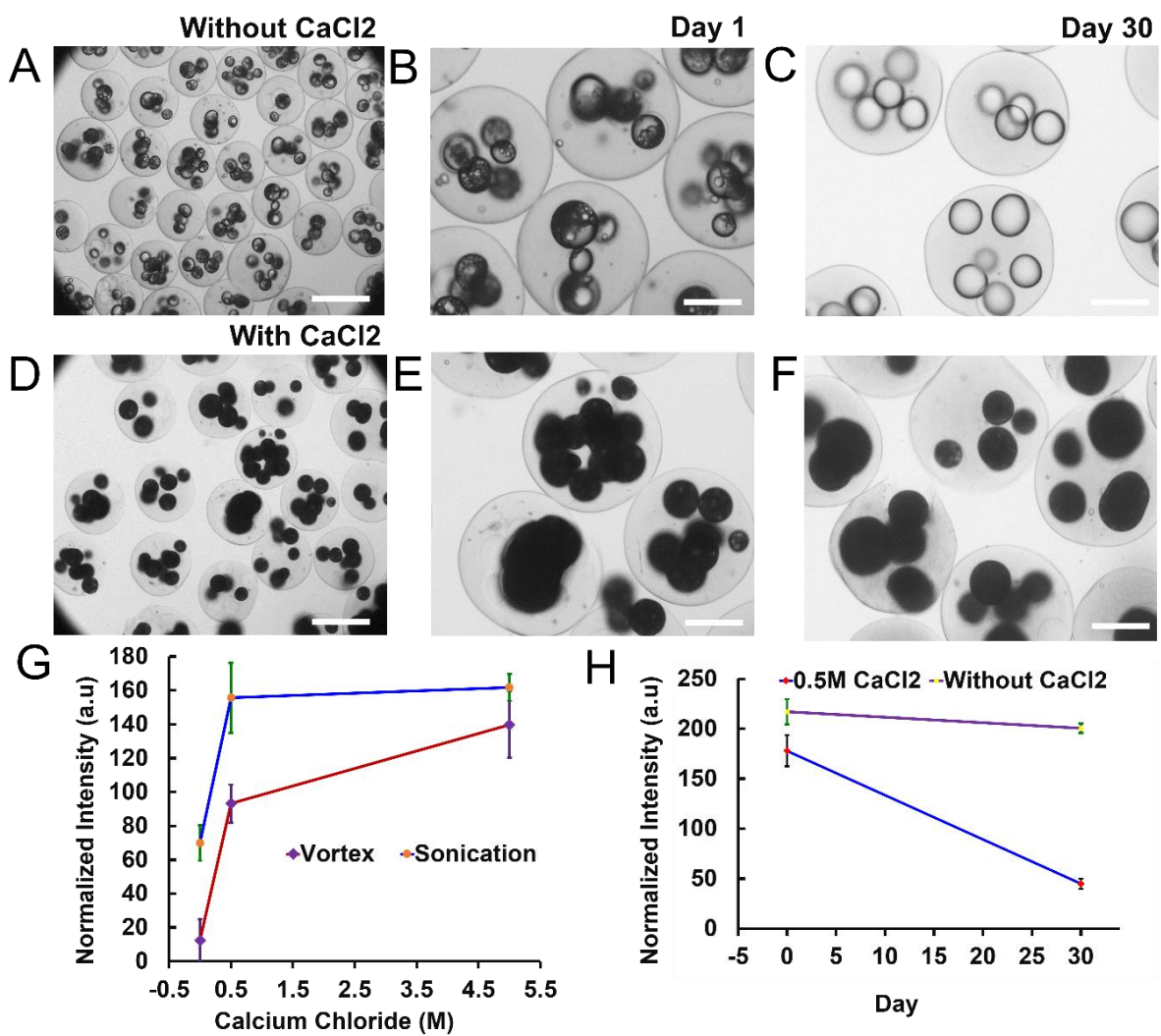

**Fig. 3 Balancing the osmotic pressure of the inner and outer aqueous phases in water/oil/water emulsions by adding calcium chloride ($CaCl_2$) to the internal phase.** A–C, no addition of $CaCl_2$ in the aqueous phase led to primary emulsions being unstable on day 1, with complete release of molecules by 1 month. D–F, in the presence of $CaCl_2$, primary emulsions remain stable (black) on day 1 of formation and continue to remain stable even after 1 month. G, the fluorescein intensity within the primary emulsion being generated using sonication or vortexing was compared, with the latter method resulting in more stable emulsions. H, fluorescein intensity was monitored with and without the presence of $CaCl_2$ in the primary emulsion. The samples which did not contain $CaCl_2$ had a reduced intensity by around 3.6 times.

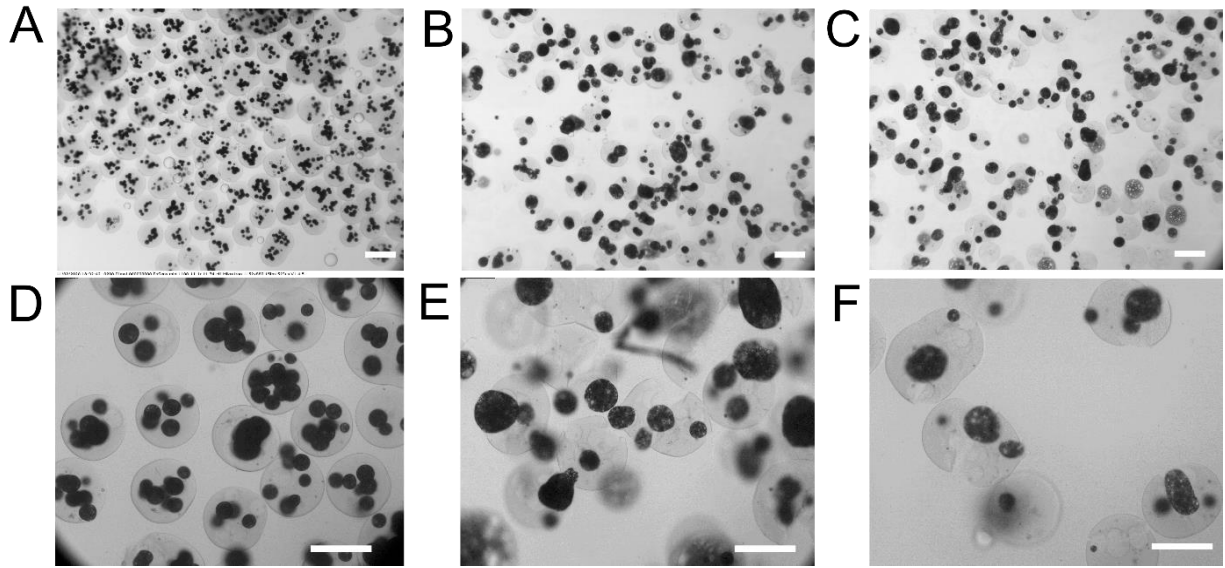

**Fig. 4 De-emulsification and washing of microcapsules with calcium chloride CaCl2, significantly enhanced the elastic modulus and textural properties of the hydrogels.** A & D, the gel hardness increased with the addition of $CaCl_2$ at an optimized concentration of 0.5 M and the microcapsules did not swell, break, or release their core encapsulated molecules. B, C, E & F, the microcapsules which were not washed with $CaCl_2$ swelled and broke due to the osmotic pressure imbalance, releasing core encapsulated molecules and causing the loss of protection and regulation of the active ingredients.

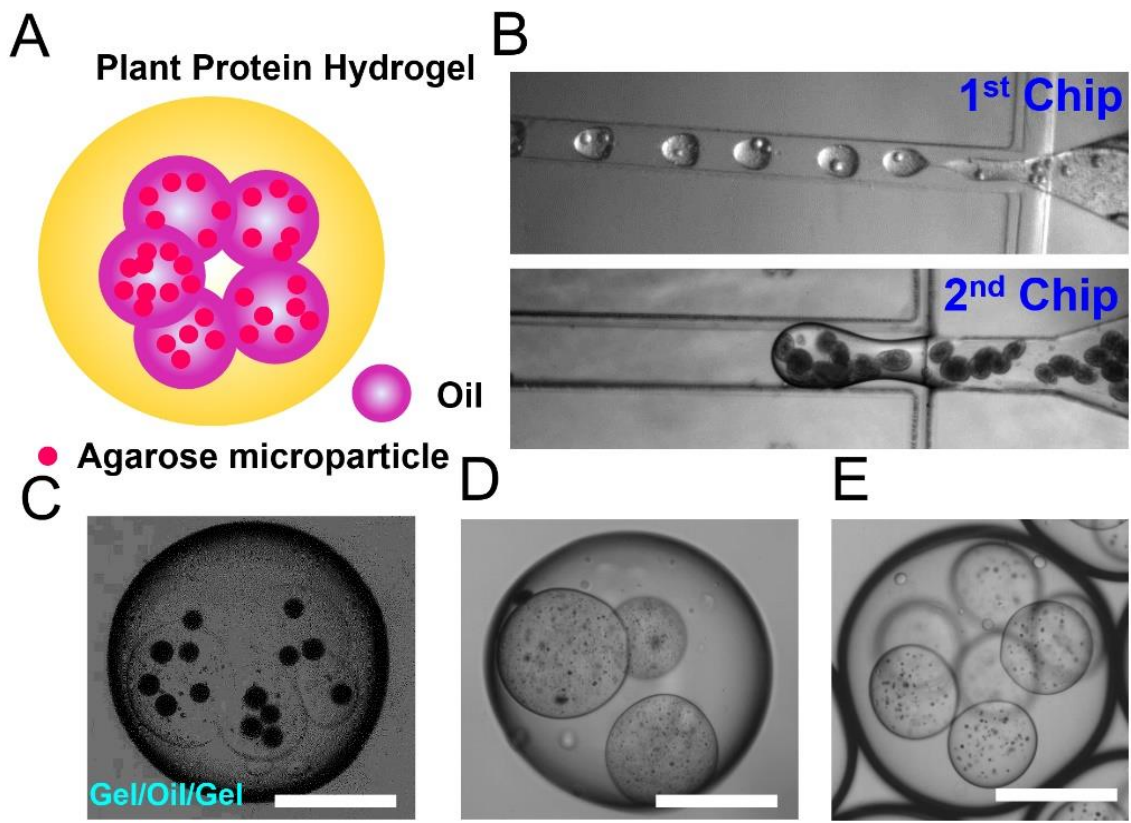

**Fig. 5. Multifunctional encapsulation of plant protein microcapsules.** A, particle/oil/particle formation with encapsulated crystallized vitamin in oil. B, the tandem emulsification system was used to fabricate protein microcapsule encapsulated food-grade agarose microparticles. Hydrophilic active ingredients, such as fluorescein, vitamin C, vitamin $B_2$, and iron, present in the aqueous inner phase were encapsulated in agarose microparticles. The middle phase was coconut oil with food-grade surfactant polyglycerol polyricinoleate (Chip 1). C, agarose microparticles in oil were encapsulated by the protein phase. D & E, the protein microcapsules were generated by encapsulating fat-soluble ingredients, such as fragrances, essential oils, and vitamin E.

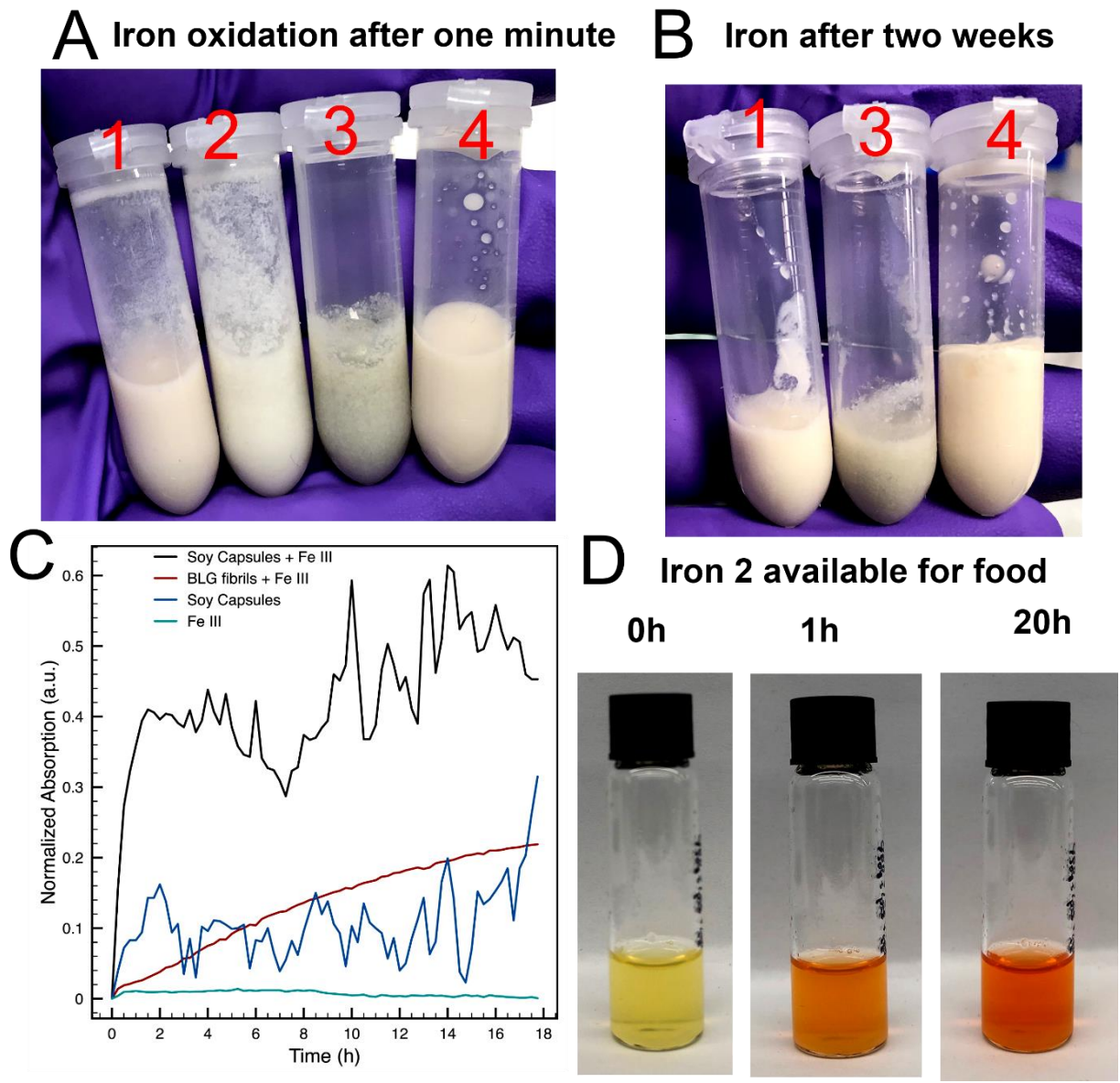

**Fig. 6 Microcapsule encapsulation prevents interactions between the encapsulated iron and oxidizing chemicals present in food.** A & B, unencapsulated iron interacted with polyphenols in banana milk, changing the color of the milk to purple-blue (tube 3; tube 2 is low-concentration iron). The microcapsule-encapsulated iron in banana milk showed no color change after 2 weeks (Tube 4). For reference, the control sample is shown in tube 1. The color change was observed at 1 min and after 2 weeks for all samples. C & D, protein fibrils added to ferric chloride showed a direct relationship between Fe(II) ions detected and the number of fibrils added. Fe(II) ion concentration was measured by the absorbance of the Fe(II)–phenanthroline complex at 512 nm. In the images to the right of the plot, greater color intensity corresponds to greater conversion of Fe(III) ions to Fe(II) ions.

# Hierarchical Plant Protein Microcapsules for Hydrophilic and Hydrophobic Cargo molecules

Ngoc-Duy Dinh[a,b], Marc Rodríguez Garcia[a,d], Zenon Toprakcioglu[a], Yi Shen[a,c] and Tuomas Knowles[a]

[a]Yusuf Hamied Department of Chemistry, University of Cambridge, Cambridge CB2 1EW, UK

[b]Department of Biomedical Engineering, The Chinese University of Hong Kong, Hong Kong SAR

[c]School of Biomedical Engineering, The University of Sydney, Australia

[d]Xampla Ltd, Cambridge, UK

**This PDF file includes:**

Tables  S1, S2
Figures  S1 to S4
Movie  S1 to S3

| Biodegradation (%) | | | | | |
|---|---|---|---|---|---|
| Test series | ThOD (mg/g test item) | Net $O_2$ (mg/l) | AVG | SD | REL |
| Cellulose (reference) | 1135 | 93.6 | 82.4 | 1.4 | 100.0 |
| Plant protein microcapsule | 1409 | 123.1 | 87.5 | 3.2 | 106.1 |

**Table. S1 Total organic carbon content, net $O_2$ production, and biodegradation percentage of cellulose and the tested microcapsule.**

ThOD: theoretical oxygen demand, AVG: average biodegradation, SD: standard deviation, REL: relative biodegradation

| Biodegradation (%) | | | | | |
|---|---|---|---|---|---|
| Test series | TOC (%) | Net $CO_2$ (mg/l) | AVG | SD | REL |
| Cellulose (reference) | 43.6 | 33.7 | 84.2 | 4.0 | 100.1 |
| Plant protein microcapsule | 49.3 | 37.2 | 82.4 | 2.1 | 98.0 |

**Table. S2. Total organic carbon content, net $CO_2$ production, and biodegradation percentage of cellulose and the tested microcapsule.**

TOC: total organic carbone, AVG: average biodegradation, SD: standard deviation, REL: relative biodegradation

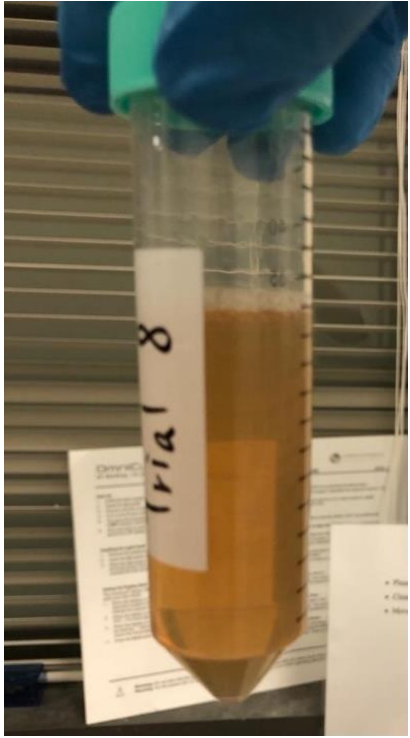

**Fig. S1. Translucent pea protein isolate in aqueous acetic acid after ultrasonication.**

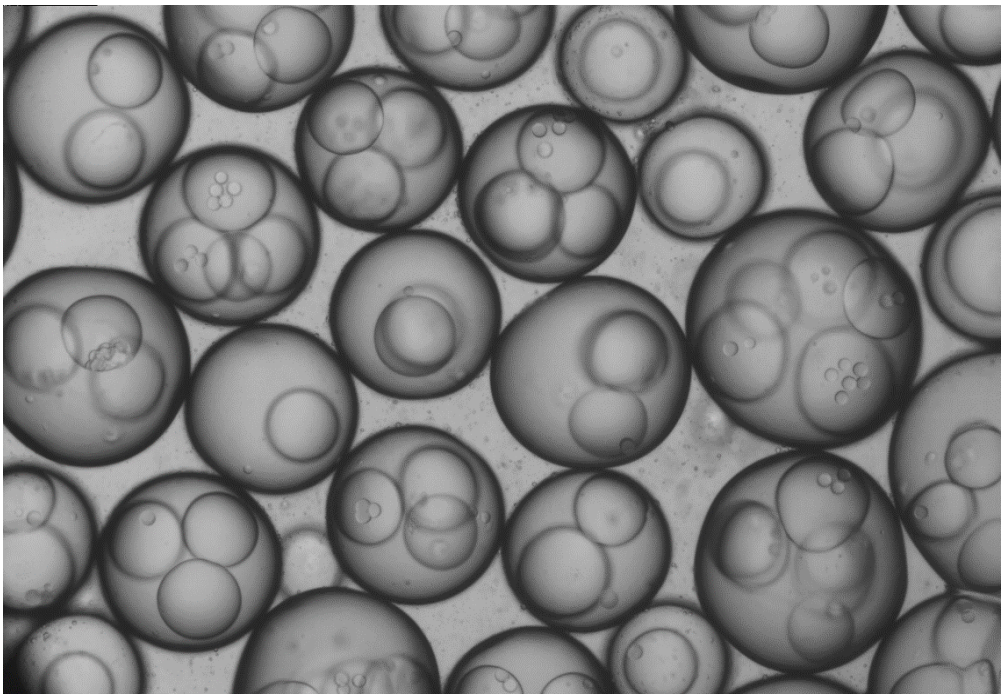

**Fig. S2. Multifunctional encapsulation with particle/oil/particle encapsulating crystallized vitamin in oil.**

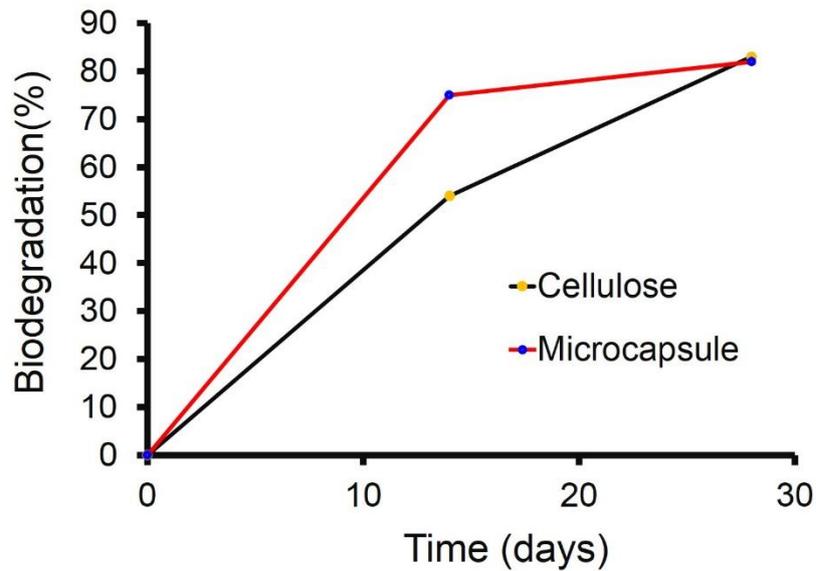

**Fig. S3.** Degradation of the tested microcapsules and reference material (cellulose) after 28 days (based on $CO_2$ production).

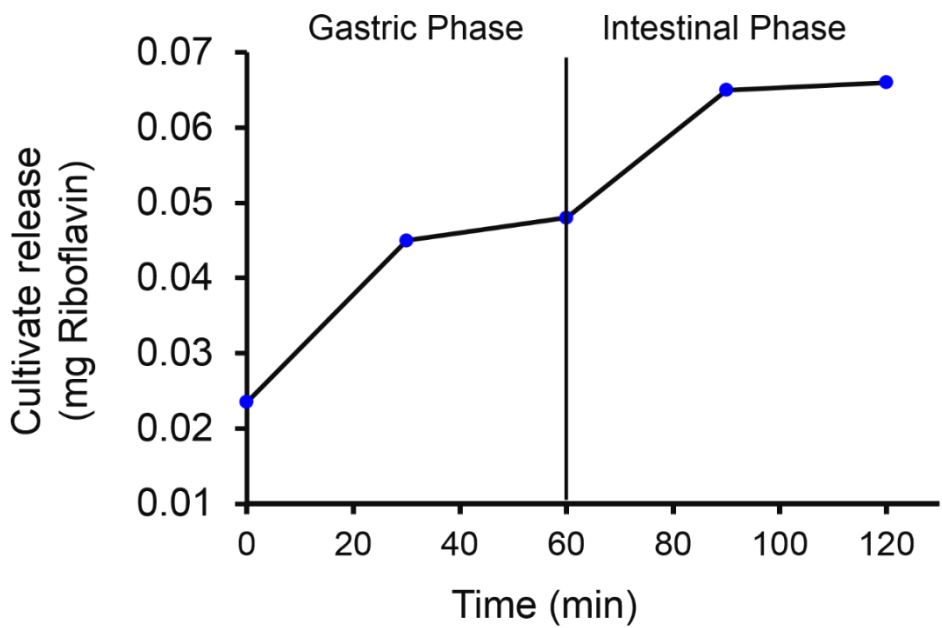

**Fig. S4.** Cumulative release of riboflavin by simulating digestive enzymes over 120 mins.

Movie S1: Chip 1. The fluorescein solution is sonicated in refined coconut oil with food-grade surfactant polyglycerol polyricinoleate to create the primary water-in-oil emulsions, which are encapsulated by plant proteins in the aqueous phase.

Movie S2: Chip 2. The double emulsions are formed in Chip 1 and continuously transferred to Chip 2, wherein oil-in-water droplets (proteins) are encapsulated by the continuous oil phase resulting in triple emulsions water-in-oil-in-water-in-oil.

Movie S3. Agarose microparticles are fabricated via droplet microfluidics.